\documentclass[12pt]{article}

\usepackage[T1]{fontenc}
\usepackage{textcomp}
\usepackage{mathptmx}

\newcommand{\ket}[1]{|#1\rangle}
\newcommand{\bra}[1]{\langle #1 |}

\title{Narrow basis angle doubles secret key in the BB84 protocol}
\author{Ryutaroh Matsumoto\\
Department of Communications and Integrated Systems\\
Tokyo Institute of Technology, 152-8550 Japan\\
Email: ryutaroh@rmatsumoto.org\\
and\\
Shun Watanabe\\
Department of Information Science and Intelligent Systems\\
Tokushima University, 770-8506 Japan\\
Email: shun-wata@is.tokushima-u.ac.jp}
\date{June 2009}

\begin{document}
\maketitle
\begin{abstract}
\noindent
We consider a modified version of 
the BB84 quantum key distribution protocol
in which the angle between two different bases are less than
$\pi/4$. We show that the channel parameter estimate becomes the
same as the original protocol with sufficiently many transmitted
qubits. On the other hand, the statistical correlation between
bits transmitted in one basis and those received in the other basis
becomes stronger as the angle between two bases becomes narrower.
If the angle is very small, the statistical correlation between
bits transmitted in one basis and those received in the other basis
is as strong as those received in the same basis as transmitting
basis, which means that the modified protocol can generate almost twice
as long secret key as the original protocol,
provided that Alice and Bob choose two different bases with almost
the same probability.
We also point out that the reverse reconciliation often gives 
different amount of 
secret key  to the direct reconciliation over Pauli channels
with our modified protocol.\\
PACS number: 03.67.Dd
\end{abstract}

\section{Introduction}
The Bennett-Brassard 1984 protocol (BB84 protocol)
\cite{bennett84}
is one of the most-known protocols for
quantum key distribution (QKD).
In this protocol,
the sender, Alice, sends qubits in one of four quantum states,
represented by quantum state vectors $\ket{0}$, $\ket{1}$,
$\ket{+} = (\ket{0}+\ket{1})/\sqrt{2}$,
$\ket{-} = (-\ket{0}+\ket{1})/\sqrt{2}$,
where $\{ \ket{0}$, $\ket{1}\}$ forms an orthonormal basis.
Then the receiver, Bob, measures them
with either $\{\ket{0}$, $\ket{1}\}$ or $\{\ket{+}$, $\ket{-}\}$
basis.
After that,
Alice publicly announces to which $\{\ket{0}$, $\ket{1}\}$ or $\{\ket{+}$, $\ket{-}\}$ basis each qubit belongs.
Bob discard the measurement outcomes whose bases do not contain
the transmitted states.
We call such measurement \emph{mismatched measurement}
in this paper.
After that, Alice and Bob perform the information reconciliation
and the privacy amplification to obtain the same secret key
as described in Ref.~\cite{shor00}.
In this standard protocol,
we have $\ket{+} = \cos \theta \ket{0} + \sin \theta \ket{1}$
and $\ket{-} = -\sin\theta\ket{0} + \cos\theta\ket{1}$ with
$\theta = \pi / 4$. In this paper we shall call $\theta$ as
the angle between two bases.

As far as the authors know,
there is no literature that shows
a merit of using smaller values of $\theta$ in the BB84 protocol,
while Tamaki et~al.\ \cite{tamaki03}
quantitatively demonstrated the merit of adjusting the angle
between two different quantum states in the Bennett 1992 (B92) protocol
\cite{bennett92qkd}.
A possible reason for the absence of consideration of narrower angle
$\theta < \pi/4$
is that the narrower angle  makes it difficult to obtain
a meaningful lower bound on the amount of secret key
by the conventional channel parameter estimation as described in Section \ref{sec22}. This difficulty leads us to use the accurate channel parameter estimation method \cite{watanabe08} for the BB84 protocol with narrower angle.
We shall show that over any quantum channel between Alice and Bob,
including Pauli channels,
we can obtain almost the same amount of secret key from
mismatched measurement outcomes when the angle between two bases
is sufficiently narrow,
while obtaining asymptotically the same amount of key per
transmitted qubit
from matched measurement outcomes,
by using the accurate estimation method.
We note that we already considered to obtain secret key from
mismatched measurement outcomes in Ref.\ \cite{matsumoto08}.
However Ref.\ \cite{matsumoto08} was not so useful because
we cannot obtain secret key if the channel is a Pauli one.

On the other hand, the amount of secret key is the same in
the direct and reverse reconciliations in the standard BB84
protocol over Pauli channels \cite{shor00}, even
if we use the accurate channel parameter estimation \cite{watanabe08}.
In contrast to this,
we also point out that the reverse reconciliation \cite{boileau05,maurer93}
often gives 
different amount of 
secret key  to the direct reconciliation over Pauli channels
with our modified protocol.

This paper is organized as follows:
Section~\ref{sec2} presents a modified version of the BB84 protocol,
its security, and its performance analysis.
Section~\ref{sec3}
gives concluding remarks.

\section{Protocol}\label{sec2}
\subsection{Outline of the protocol}\label{sec21}
In this section,
we shall show a variant of the BB84 protocol
that tries to extract secret key from mismatched measurement outcomes.
Section \ref{sec21} describes an outline  of the protocol,
Section \ref{sec22} derives the amount of secret key,
and Section \ref{sec24} considers the reverse reconciliation.
We define the matrices $X$ and $Z$ representing
the bit error and the phase error, respectively, as
\begin{eqnarray*}
&&X\ket{0} = \ket{1},\quad X\ket{1}=\ket{0},\\
&&Z\ket{+} = \ket{-},\quad Z\ket{-}=\ket{+},
\end{eqnarray*}
and $Y= i XZ$.
We also fix $0 < \theta \leq \pi/4$ and define
\begin{eqnarray*}
\ket{+_\theta} &=& \cos \theta \ket{0} + \sin \theta \ket{1},\\
\ket{-_\theta} &=& -\sin\theta\ket{0} + \cos\theta\ket{1}.
\end{eqnarray*}

\begin{enumerate}
\item\label{l1}
Alice makes a random qubit sequence according to
the i.i.d.\ uniform distribution on $\{\ket{0}$, $\ket{1}$,
$\ket{+_\theta}$, $\ket{-_\theta}\}$ and sends it to Bob.
\item\label{l2}
Bob chooses the $\{\ket{0}$, $\ket{1}\}$ basis or $\{\ket{+_\theta}$, $\ket{-_\theta}\}$
basis uniformly randomly for each received qubit and measures it
by the chosen basis.
\item\label{l4}
Alice publicly announces which basis $\{\ket{0}$, $\ket{1}\}$ or
$\{\ket{+_\theta}$, $\ket{-_\theta}\}$ each transmitted qubit belongs to.
Bob also publicly announces which basis was used for measurement
of each qubit.
\item\label{l45}
Suppose that there are $2n$ qubits transmitted in the $\{\ket{0}$, $\ket{1}\}$
basis and measured with the $\{\ket{+_\theta}$, $\ket{-_\theta}\}$ basis by Bob.
Index those qubits by $1$, \ldots, $2n$.
Define the bit $x_i = 0$ if Alice's $i$-th qubit was $\ket{0}$,
and $x_i=1$ otherwise.
Define the bit $y_i = 0$ if Bob's measurement outcome
for $i$-th qubit was $\ket{+_\theta}$,
and $y_i=1$ otherwise.
\item\label{l5}
Suppose also that there are $2n'$ qubits transmitted in the $\{\ket{0}$, $\ket{1}\}$
basis and measured with the $\{\ket{0}$, $\ket{1}\}$ basis by Bob.
Index those qubits by $1$, \ldots, $2n'$.
Define the bit $a_i = 0$ if Alice's $i$-th qubit was $\ket{0}$,
and $a_i=1$ otherwise.
Define the bit $b_i = 0$ if Bob's measurement outcome
for $i$-th qubit was $\ket{0}$,
and $b_i=1$ otherwise.
\item\label{l6}
Suppose also that there are $2n''$ qubits transmitted in the $\{\ket{+_\theta}$, $\ket{-_\theta}\}$
basis and measured with the $\{\ket{+_\theta}$, $\ket{-_\theta}\}$ basis by Bob.
Index those qubits by $1$, \ldots, $2n''$.
Define the bit $\alpha_i = 0$ if Alice's $i$-th qubit was $\ket{+_\theta}$,
and $\alpha_i=1$ otherwise.
Define the bit $\beta_i = 0$ if Bob's measurement outcome
for $i$-th qubit was $\ket{+_\theta}$,
and $\beta_i=1$ otherwise.
\item For each combination of the transmission and the reception bases,
Alice and Bob publicly announce the half of transmitted qubits and
measurement outcomes. They conduct the channel parameter estimation
described in Section \ref{sec22}.
We also define
\[
q_1 = \frac{|\{i\in S \mid x_i \neq y_i \}|}{|S|},\quad
q_2 = \frac{|\{i\in S' \mid a_i \neq b_i \}|}{|S'|},
\]
where $S$ and $S'$ are the set of indices that are announced for
channel parameter estimation.
\item\label{l9}
Alice and Bob decide\footnote{One can also use
the Slepian-Wolf code used in Ref.\ \cite{watanabe08}.} a linear code $C_1$ of length $n$
such that its decoding error probability is sufficiently small
over all the binary symmetric channel whose crossover probability
is close to $q_1$.
Let $H_1$ be a parity check matrix for $C_1$,
$\vec{x}$ be Alice's remaining (not announced) bits among $x_i$'s,
and $\vec{y}$ be Bob's remaining bits among $y_i$'s.
\item\label{l10}
Alice publicly announces the syndrome $H_1\vec{x}$.
\item\label{l12}
Bob computes the error vector $\vec{e}$ such that
$H_1\vec{e}=H_1\vec{y}-H_1\vec{x}$ by the decoding
algorithm for $C_1$. With high probability $\vec{y}-\vec{e} = \vec{x}$.
\item\label{l13}
Alice chooses a subspace $C_2 \subset C_1$ with $\dim C_2 = n(1-S(X|E)+\epsilon)$
uniformly randomly, where $\epsilon > 0$ and
$S(X|E)$ denotes the conditional von Neumann
entropy of Alice's bit $x_i$ given the quantum state of the environment
$E$ as defined in Refs.\ \cite{rennerphd,renner05},
which can be regarded as the eavesdropper Eve's ambiguity on Alice's bit $x_i$.
After that she publicly announces her choice of $C_2$.
The final shared secret key is the coset $\vec{x}+C_2$.
\end{enumerate}
Provided that $\epsilon > 0$, the privacy amplification theorem
with quantum eavesdropper's memory
\cite{rennerphd,renner05} guarantees that that the final
key $\vec{x}+C_2$
becomes secure in the sense of Refs.\ \cite{rennerphd,renner05}
as $n\rightarrow
\infty$, which roughly means that
the final key and the quantum
state of the environment become statistically independent and
that the final key has an almost uniform distribution
on the set $C_1/C_2$.
We shall consider the amount of secret key obtained by the above
protocol in Section \ref{sec22}.

\subsection{Amount of secret key}\label{sec22}
We shall use the accurate channel parameter estimation \cite{watanabe08},
which gives asymptotically more secret key than the conventional estimation.
This procedure is as follows:
We do not make any assumption on the quantum channel
between Alice and Bob, so the channel is specified by
12 real parameters. 
For 16 pairs $(\ket{u}$, $\ket{v}) \in \{\ket{0}$,
$\ket{1}$, $\ket{+_\theta}$, $\ket{-_\theta}\}^2$,
we record the 16 relative frequencies of the events in which
$\ket{u}$ is transmitted
and $\ket{v}$ is observed as the measurement outcome,
which enable us to estimate 6 out of 12 channel parameters.
After estimating the part of parameters,
we use the minimum of $S(X|E)$ over all the possible quantum channels,
that is, we use the worst case estimate of $S(X|E)$ of quantum channels
giving the 16 recorded relative frequencies,
as done in the conventional estimation \cite{gottesman03,rennerphd,renner05}.
The set of estimatable parameters with $0 < \theta < \pi / 4$
is the same as $\theta = \pi/4$. The reason is as follows:
Since the linear space spanned by
$\{\ket{0}\bra{0}$, $\ket{1}\bra{1}$,
$\ket{+_\theta}\bra{+_\theta}$, $\ket{-_\theta}\bra{-_\theta}\}$
is the same for all $0<\theta\leq \pi/4$
and the expectation of the relative frequency of sending
$\ket{u}$ and observing $\ket{v}$ is proportional to
$\mathrm{Tr}[\Lambda(\ket{u}\bra{u})\ket{v}\bra{v}]$
for any quantum channel $\Lambda$,
there always exists a one-to-one linear relation that
translates the set of 16 relative frequencies with
$\theta < \pi/4$ to that with $\theta = \pi/4$.
Therefore, the estimate of
the worst case $S(X|E)$ does not depend on the value of $\theta$.
This means that the amount of secret key from matched
measurement outcomes remains asymptotically the same even if
we use $\theta$ narrower than $\pi/4$.

We cannot use a straightforward generalization of the conventional channel
parameter estimation, that is to record two relative frequencies of the
event (a) in which $\ket{0}$ is sent and $\ket{1}$ is observed or
$\ket{1}$ is sent and $\ket{0}$ is observed, and the event (b) in which
$\ket{+_\theta}$ is sent and $\ket{-_\theta}$ is observed or
$\ket{-_\theta}$ is sent and $\ket{+_\theta}$ is observed.
The reason of unavailability of the conventional channel parameter
estimation
is as follows:
We cannot estimate the parameters of the Pauli channel
that is obtained as the partial twirling\footnote{%
See also Eq.\ (12) of Ref.\ \cite{hamada-epp},
in which the partial twirling is called the discrete twirling.} \cite{bennett96}
of the actual quantum channel, because the relative frequency of
the event (b) also depends\footnote{%
The relative frequency of the event (a) is independent
of the non-diagonal elements in the Choi
matrix.} on the non-diagonal elements in the Choi
matrix \cite{choi75} 
of the actual quantum channel with respect to the Bell basis as well as
the diagonal elements unless $\theta = \pi/4$,
and the 4 diagonal elements in the Choi matrix specify
the Pauli channel obtained by the partial twirling.
Since the standard technique is
to bound the required dimension of $C_2$ in Step \ref{l13}
over the actual channel
from above by the required $\dim C_2$ over its partially twirled channel,
the inability to estimate the partially twirled channel
prevents us from  obtaining a useful upper bound on $\dim C_2$
of the actual channel.
Thus, it is difficult to ensure that the worst case estimate
of $\dim C_2$ is independent of $\theta$
by the above generalization of the conventional channel
parameter estimation, and
we have to use the 16 relative frequencies
to bound $\dim C_2$ from above.
Note that Tamaki et~al.\ \cite{tamaki03} already
observed similar dependence of the worst case estimate
on the angle between two quantum states in the B92 protocol \cite{bennett92qkd}.

The amount of secret key is \cite{rennerphd,renner05}
\[ 
S(X|E) - h(q_1) 
\] 
from single bit $x_i$ not announced for channel parameter estimation,
while this amount is
\begin{equation}
S(X|E) - h(q_2) \label{rate:match}
\end{equation}
from $a_i$, where $h()$ denotes the binary
entropy function. Since $q_1 \rightarrow q_2$ as $\theta \rightarrow
0$ and $h()$ is a continuous function,
we conclude that we can obtain almost the same amount of
secret key from $x_i$ as $a_i$.

\subsection{Reverse reconciliation}\label{sec24}
The reverse reconciliation \cite{boileau05,maurer93}
is the method of reconciliation
in which Bob publicly announces the syndrome $H_1\vec{y}$ in Step
\ref{l10} instead of Alice, Alice computes $\vec{y}$ in Step \ref{l12}
instead of Bob, and the final key is generated from $\vec{y}$.
The standard way of reconciliation \cite{shor00}
is called the direct reconciliation.
In order to give a simpler
exposition of the main contribution, we have restricted ourselves to
the direct reconciliation up to this point.
In this subsection we shall consider the
reverse reconciliation and point out that the amount of secret key
is often different in the reverse reconciliation to the direct
one over a Pauli channel.

We can also use the same parity check matrix $H_1$ in Step \ref{l9}
since $\vec{x}$ can be regarded as the output of the binary
symmetric channel with crossover probability $q_1$ with input $\vec{y}$.
We have to change $\dim C_2$ in Step \ref{l13} to
$\dim C_2 = n(1-S(Y|E)+\epsilon)$,
where
$S(Y|E)$ denotes the conditional von Neumann
entropy of Bob's bit $y_i$ given the quantum state of the environment
$E$.
We have to compute the minimum value of $S(Y|E)$
over quantum channels that give the recorded relative frequencies.

Hereafter we assume that the channel between Alice and Bob is
a Pauli channel that sends a qubit density matrix $\rho$ to
\[
\Gamma(\rho) = (1-r_X - r_Y - r_Z) \rho + r_X X\rho X + r_Y Y\rho Y + r_Z Z \rho Z,
\]
instead of a general qubit channel that is not necessarily a Pauli one.
We define
\[p_X = r_X + r_Y, \quad p_Z = r_Z + r_Y.
\]
It is well-known that the worst case $S(X|E)$ is $1-h(p_Z)$
\cite{gottesman03,rennerphd,renner05}.

In the evaluation of the worst case $S(Y|E)$,
Bob's bit $Y$
can be regarded as the measurement outcome
in the $\{\ket{0}$, $\ket{1}\}$ basis on the output
of the unitary channel rotating $\ket{+_\theta}$ to
$\ket{0}$ and $\ket{-_\theta}$ to $\ket{1}$,
connected to the actual channel.
The Pauli channel followed by a rotation is a unital channel,
which outputs the completely mixed state if the input is
completely mixed.
This observation enables us to apply the formula for the worst case
$S(Y|E)$ over unital channels given in Proposition 2 and Remark 6
of Ref.\ \cite{watanabe08}, which gives
\[
S(Y|E)
 = 1 - h(p_X) - h(p_Z)
+ h\left(\frac{1 + \sqrt{(1-2p_X)^2\cos^2 2\theta+(1-2p_Z)^2 \sin^2 2\theta}}{2}\right).
\]
We can see that
$S(Y|E) \rightarrow 1-h(p_Z)$ as $\theta \rightarrow 0$ and
$S(Y|E) \rightarrow 1-h(p_X)$ as $\theta \rightarrow \pi/4$,
which confirm our intuition.
Observe also that generally $S(X|E) \neq S(Y|E)$ when
$p_X \neq p_Z$.

By using a similar idea,
we can obtain Eve's ambiguity on Alice's bit $\alpha_i$ that is
transmitted in the $\{\ket{+_\theta}$, $\ket{-_\theta}\}$ basis.
By the continuity of the von Neumann entropy, we can also see that
the amount of secret key from $\alpha_i$
converges to Eq.\ (\ref{rate:match}) obtained from
$a_i$
as $n \rightarrow \infty$ and $\theta \rightarrow 0$.
Therefore, the conclusion in Section \ref{sec22} also holds
for qubits transmitted by the $\{\ket{+_\theta}$, $\ket{-_\theta}\}$ basis.

\section{Concluding remarks}\label{sec3}
We have shown that from mismatched measurement outcomes
we can obtain as much secret key per transmitted qubit
as matched measurement outcomes
over any channels
if we make the angle between two bases sufficiently narrow.
The same conclusion holds for the six-state protocol
\cite{bruss98b} and the variants of the standard
BB84 protocols with the noisy preprocessing \cite{rennerphd,renner05},
and the advantage distillation \cite{gottesman03,watanabe07}.
We have also pointed out that the reverse reconciliation often gives 
different amount of
secret key to the direct reconciliation over Pauli channels
with our modified protocol,
which is contrasting to the standard BB84 protocol \cite{shor00},
and that there is difficulty to use the conventional channel
parameter estimation if the angle between two bases is narrower
than $\pi/4$.

The advantage of the proposed protocol is that
we can obtain $1-h(p_X)-h(p_Z)$ bits of secret key per
single qubit that is not used for channel parameter estimation.
The same advantage is also realized when we
decrease the ratio of the number of transmitted qubits
in the $\{\ket{+}$, $\ket{-}\}$ basis to that in the
$\{\ket{0}$, $\ket{1}\}$ basis \cite{hayashi09,lo05}.
Although the proposed method, the method in Refs.\ \cite{hayashi09,lo05},
and their combination have exactly the same performance
in the asymptotic limit of infinitely many qubits,
they may have different performances in the finite number of
qubits. The identification of the best method among these
three methods in the finite setting
is a future research agenda. This identification might be analytically
difficult as stated in
the introduction of Ref.\ \cite{hayashi09}.


\end{document}